\documentstyle[psfig,multicol,aps,epsf]{revtex}
\begin{document}
\twocolumn[\hsize\textwidth\columnwidth\hsize\csname
 @twocolumnfalse\endcsname
\vspace*{-1.5cm}
\begin{center}
{\bf Published in {\em Phycical Review Letters} (Phys. Rev. Lett. 88(1), 018701-1 (2002)).} 
\end{center}
\vspace*{0.5cm}
\title{A Complexity View of Rainfall}
\author{Ole Peters$^1$, Christopher Hertlein$^{1,2}$, and Kim
Christensen$^{1,*}$}
\pagestyle{myheadings}
\address{{$^1$}Blackett Laboratory, Imperial College, Prince Consort Road,
London SW7 2BW, United Kingdom \\
{$^2$}Fakult\"{a}t f\"{u}r Physik,
Albert-Ludwigs-Universit\"{a}t
Freiburg, Hermann-Herder-Stra{\ss}e, Westbau, D-79104 Freiburg, Germany}

\vspace*{-0.25cm}
\maketitle
\date{\today}
\begin{abstract}
We show that rain events are analogous to a variety of nonequilibrium
relaxation processes in Nature such as earthquakes and avalanches.
Analysis of high-resolution rain data reveals that power laws describe
the number of rain events versus size and number of droughts versus duration.
In addition, the accumulated water column displays scale-less fluctuations.
These statistical properties are the fingerprints of a self-organized critical
process and may serve as a benchmark for models of precipitation and
atmospheric processes.
\end{abstract}
\vskip2pc]

Rainfall and rainfall-related quantities have been recorded for centuries
\cite{1,2}. All these measurements, however, have the disadvantage of low
temporal resolution and low sensitivity. The rain measurements are based
on the simple idea of collecting rain in a container and measuring the
amount of water after a certain time. The time intervals between readings
are typically hours or days. Even with the most sophisticated of these 
conventional methods, the fine details of rain events cannot be captured
at all and very light rain might not be recorded due to evaporation or
insufficient sensitivity of the instrument, making it impossible to
address questions regarding single rain events.

Recently, high-resolution data have been collected with a compact vertically 
pointing Doppler radar MRR-2, developed by METEK\cite{3}. The instrument
is operated by the Max-Planck-Institute for Meteorology, Hamburg, Germany
at the Baltic coast Zingst (54$^{\circ}$43'N 12$^{\circ}$67'E) under the
Precipitation and Evaporation Project (PEP) in BALTEX \cite{4}. Rain rate,
liquid water content, and drop size distribution were obtained from the 
radar Doppler spectra, based on a method described by Atlas \cite{5,6,7}.
At vertical incidence, the Doppler shift can be identified with the
droplet fall velocity. As, in the atmosphere, larger drops fall faster
than smaller drops, spectral bins can be attributed to corresponding 
drop sizes.
For a given size, the scattering cross section of
the droplets can be calculated by Mie theory \cite{Mie1908}. This
yields the number density of drops which is proportional to the
spectral power divided by the corresponding cross section.
The rain rate $q(t) = \sum_i n_i V_i v_i$, where $n_i$ is the
number density of drops of volume $V_i$ falling with
velocity $v_i$. The detection threshold for rain rates under the
pertinent operation parameters was $q_{min} = 0.005$ mm/h. Below this
threshold, $q(t) = 0$ by definition.

Precipitation profiles up to some thousand meters altitude can be
observed. At present, the quantitative retrieval is 
restricted to rain. Snow and hail can be identified from the form of
the Doppler spectra but have been excluded from the quantitative analysis.
The analyzed data refer to 250m above sea level and have been collected
from January to July 1999 with 1-min resolution.

\begin{figure}[h]
\vspace*{-1.75cm}
\centerline{\psfig{figure=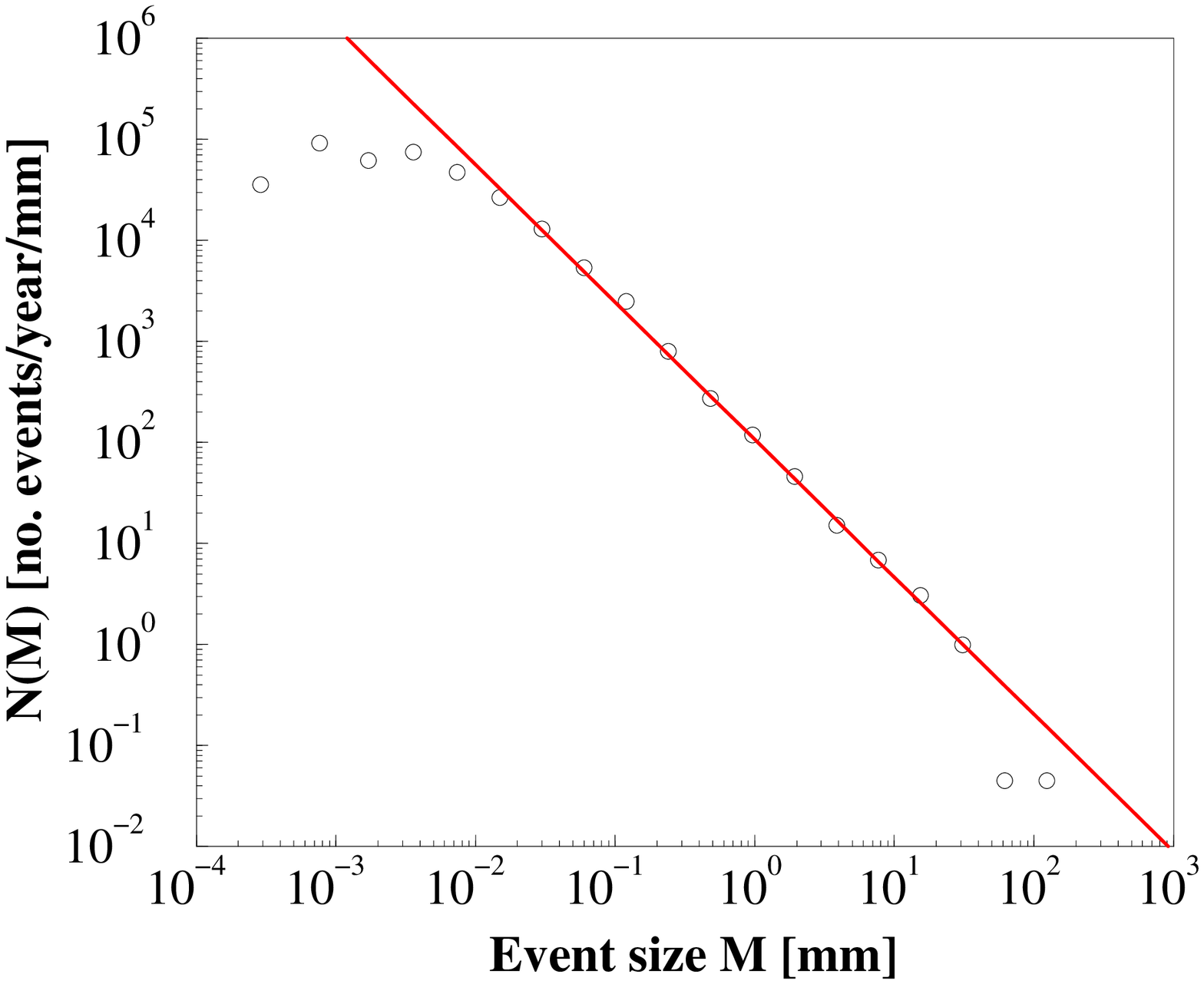,height=8cm,angle=-90}}
\caption{The number density of rain events per year $N(M)$ versus event
size $M$ (open circles) on a double logarithmic scale. A rain event
is defined as a sequence of consecutive non-zero 
rain rates (averaged over 1 min). This implies that a rain event
terminates when it stops raining for a period of at least 1 min.
The size $M$ of a rain event is the water column (volume per area) released.
Over at least three decades, the data are consistent 
with a power law $N(M) \propto M^{-1.36}$, shown as a solid line.}
\label{fig1}
\end{figure}

The processes that make a cloud release its water content are only very little 
understood. However, with the high temporal resolution of 1 min,
single rain events can be identified and characterized. Previous work
focused on the rainfall during a fixed period of time \cite{8,9,10}.
What makes the present analysis fundamentally new is the 
identification of a rain event as the basic entity. We define an event
as a sequence of successive non-zero rain rates. Sequences of zero-rain
rates in between rain events are called drought periods.
The event size is defined as the released water column in mm,
$M = \sum_t q(t) \Delta t$, where $\Delta t = 1$ min,
that is, the time-integral of the rain rate over an event.
In Fig.~\ref{fig1}, the number density of rain events per year
$N(M)$ versus event size $M$, is displayed on a double-logarithmic plot.
In a certain scaling regime, extending
over at least three decades, the number density of rain events obeys a
simple power law
\begin{equation}
N(M) \propto  M^{-1.36},
\end{equation}
represented by the solid line in Fig.~\ref{fig1}.

\begin{figure}[h]
\vspace*{-1.75cm}
\centerline{\psfig{figure=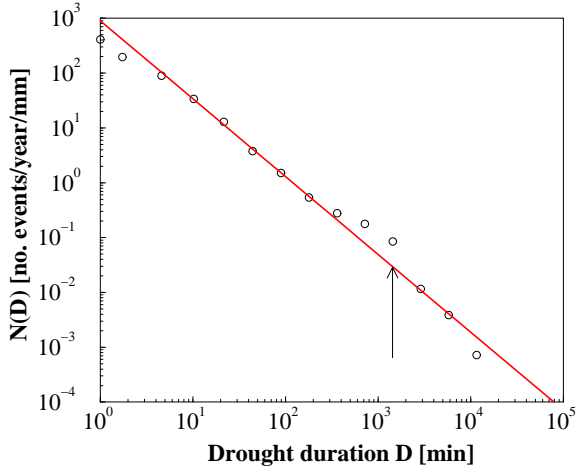,height=8cm,angle=-90}}
\caption{
The number density of droughts per year $N(D)$ versus drought duration $D$
(open circles) on a double logarithmic plot.
The drought duration is the time, measured in minutes,
between two successive rain events.
The displayed solid line
is a power law $N(D) \propto D^{-1.42}$. The arrow 
indicates a time interval of one day. The data deviate from the
power-law behavior at time intervals corresponding to about a day,
reflecting the daily meteorological cycle.
}
\label{fig2}
\end{figure}

Figure~\ref{fig2} displays the number density of inter-occurrence times
(drought durations) $N(D)$ between successive rain events.
The drought duration is power-law distributed 
\begin{equation}
N(D) \propto D^{-1.42},
\end{equation}
implying there is no typical duration of droughts.
We were not able to detect a lower or an upper cutoff this relation.
Both the lower end (1 min) and the upper 
end (two weeks) still lie within the scaling region of the power law.
The observed deviation around a period of 1 day is related to the daily
meteorological cycle.

It is compelling, that the distributions of sizes of rain events and
drought periods are simple power laws. This result could prove very
useful in relation to drought hazard assessment or flooding hazard
assessment. In order to calculate the expected number ${\bar{N}}(T)$ of
droughts with period $D > d$ in a given time period $T$, we would
have to integrate $N(D)$. Assuming, for simplicity, that the upper cutoff
diverges, $N(D) = \mbox{const} \cdot D^{-1.42}$, we
find ${\bar{N}}(T) = T \cdot N(D>d) = T \cdot \frac{\mbox{const}}{0.42} \cdot d^{-0.42}$.

\begin{figure}[h]
\vspace*{-1.75cm}
\centerline{\psfig{figure=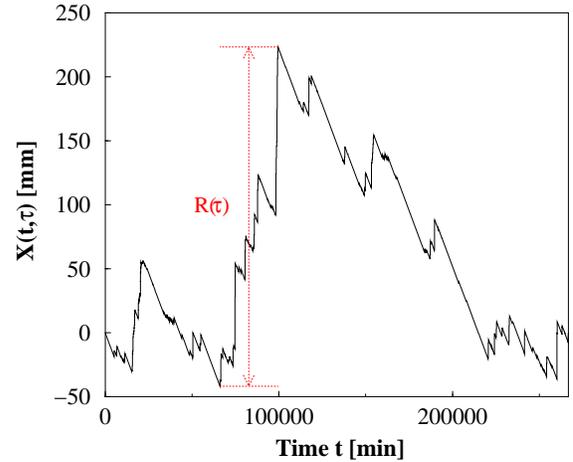,height=8cm,angle=-90}}
\caption{
Reservoir level $X(t,\tau)$ in mm for the entire record of duration
$\tau = 266,611$ min.  The parts of the curve with negative slope
correspond to dry-periods (droughts) where 
there is no rain, only the mean outflux. The parts of the graph with
positive slope are periods with rain events.
The steepness of the line measures the difference between the 
influxes and the average outflux.
The range $R(\tau)$ is indicated with a dashed line. 
}
\label{fig3}
\end{figure}

The question of having a reliable water supply is of utmost importance.
H.E. Hurst \cite{1,2} posed the following problem: How can one design
a reservoir so that it never overflows or empties?
He considered an incoming signal $q(t)$ over a time period $\tau$.
In our case, $q(t)$ is the rain rate. The actual level of water
in a reservoir (or a river ) is determined by
\begin{equation}
X(t,\tau)=\sum_{u=1}^t \left( q(u)-{\langle q \rangle}_{\tau}\right) \Delta t,
\end{equation}
where $\Delta t = 1$ min and
\begin{equation}
{\langle q \rangle}_{\tau} = \frac{1}{\tau} \sum_{t=1}^{\tau} q(t) \Delta t
\end{equation}
denotes the average 
influx in the considered time period $\tau$. The water level needed
for the reservoir never to 
empty is given by the range
\begin{equation}
R(\tau) = \max_{1 \leq t \leq \tau} X(t,\tau) - \min_{1 \leq t \leq \tau} X(t,\tau).
\end{equation}
One can now determine the 
dimensionless ratio $R(\tau)/S(\tau)$ as a function of $\tau$,
where $S(\tau)$ is the standard deviation of 
the influx $q(t)$ in the period $\tau$. For uncorrelated random events,
this ratio increases as
\begin{equation}
R(\tau)/S(\tau) \propto {\tau}^H,
\end{equation}
where the
Hurst exponent $H=1/2$. However, Hurst \cite{1,2} discovered that for 
water level fluctuation in the Nile, $H \approx 0.77$.
Figure~\ref{fig3} displays the water level $X(t,\tau)$ in a 
virtual reservoir for the rain data with $\tau = 266,611$ min.

\begin{figure}[h]
\vspace*{-1.75cm}
\centerline{\psfig{figure=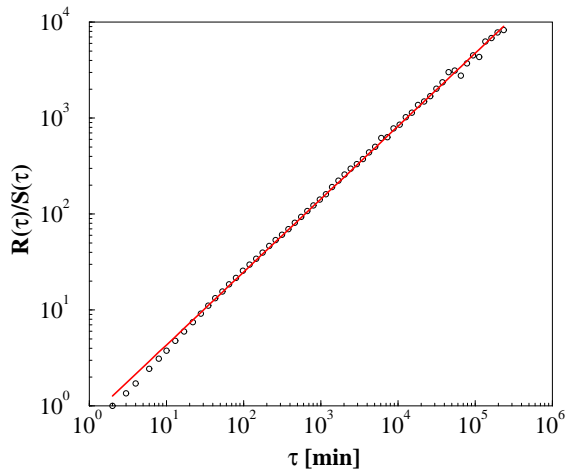,height=8cm,angle=-90}}
\caption{
Log-Log plot of values of $R(\tau)/S(\tau)$ against the corresponding
period of length $\tau$ (open circles).
The solid line is a power law
$R(\tau)/S(\tau) \propto {\tau}^H , H = 0.76$. The 
data are consistent with a power law over at least four decades.
There is a lower cutoff around $\tau \approx 10$ min but no upper
cutoff is apparent.
}
\label{fig4}
\end{figure}

Figure~\ref{fig4} demonstrates that 
$R(\tau)/S(\tau) \propto {\tau}^H, H \approx 0.76$ is obeyed over more
than four decades with
$\tau \in [10~\mbox{min}, 266,611~\mbox{min} \approx 6~\mbox{months}]$.

It is important to notice that these fluctuations are a result of the
fluctuating rain rate alone 
and imply a correlation between rain events over the whole temporal
range studied in this letter. This extends Hurst's result \cite{2},
which he found was valid in the temporal range
$\tau \in [1~\mbox{year}, 1080~\mbox{years}]$.

\begin{figure}[h]
\vspace*{-1.75cm}
\centerline{\psfig{figure=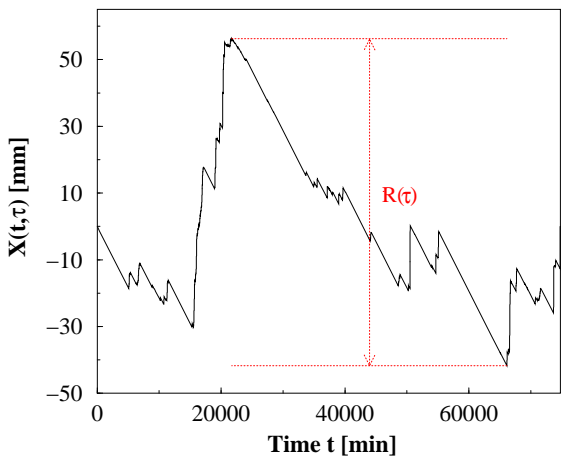,height=8cm,angle=-90}}
\caption{
Reservoir level $X(t,\tau)$ in mm for 
the initial part of the record with a duration of $\tau = 74,743$ min.
The corresponding range $R(\tau)$ is indicated with a dashed line.
Compared with Fig.~\ref{fig3}, the x-axis has been 
rescaled with a factor $b = 266,611/74,743 \approx 3.57$ while the y-axis
has been rescaled with 
a factor $b^H \approx 2.63 \approx 300/115$ demonstrating that the
reservoir level is a self-affine fractal.
}
\label{fig5}
\end{figure}

It can be illustrated directly that
the fluctuations of the reservoir 
are statistically invariant under a transformation that changes the
time scale by a factor $b$ 
and the level by a factor $b^H$ \cite{11}. 
In Fig.~\ref{fig5}, the x-axis of Fig.~\ref{fig3} has been re-scaled
with a factor $b$ and the y-axis with a factor $b^H$, and the similarity
is indeed striking! 

The power-law number density of rain events is consistent with a
self-organized critical process. The concept of self-organized
criticality \cite{12,13,14} refers to the tendency of 
non-equilibrium systems driven by slow constant energy input to
organize themselves into a critical states where all scales are relevant.
The characteristic feature of self-organized critical systems, 
even if their dynamics are incomprehensibly complex, is 
that the intermediately stored energy is eventually released in
sudden bursts with no typical size.

A well-known example of such a system is the Earth's crust.
Currents in the liquid core of the Earth drive the crust slowly
and fairly constantly. The energy deposited 
by these currents is intermediately stored in tension building
up between the tectonic plates and then suddenly released in
earthquakes. The number of earthquakes per year 
with a seismic moment $S$ exceeding $s$ is given by the
Gutenberg-Richter law \cite{15}
\begin{equation}
N(S>s) \propto s^{-B},
\end{equation}
that is, there is no typical size for earthquakes.
This suggests all the earthquakes 
have the same physical origin and that the Earth's crust is poised
in a critical state.

Avalanches in a pile of grains might also display self-organized
criticality: when grains are dropped onto a pile, one by one,
the pile ultimately reaches a stationary critical 
state in which its slope fluctuates about a constant angle of
repose, with each new grain being capable of inducing an avalanche
on any of the relevant size scales \cite{16}.

From the perspective of self-organized criticality, rain events do not
look very different from earthquakes or avalanches. If a rain shower,
regardless of its duration or intensity,
is defined as an event, the correspondence to avalanches in granular
media and avalanches in the crust of the earth is striking.
The atmosphere is the system under 
investigation and corresponds to the Earth's crust or the granular pile.
It is driven by a slow and constant energy input from the Sun.
In particular, water is evaporated from the 
oceans. The energy is stored in the form of water vapor in the atmosphere.
It is then suddenly released in bursts when the vapor condenses to water
drops. The power-law distribution of the number density of rain events versus
size is equivalent to the Gutenberg-Richter law for earthquakes and the
power-law distribution of avalanche size.
There is no constant drizzle accounting for the constant 
evaporation but rain events of a wide range of sizes.
One could imagine having a classification of rain events
according to their size just as earthquakes are classified 
according to their position on the Richter scale.

In summary, we found that simple power laws describe the number density of 
occurrence of rain events of a given size and drought periods.
Moreover, Hurst's analysis from the 1950ies on water level
fluctuations was extended by more than four decades, 
from a year down to minutes. This insight will inevitably inspire new
research into the modeling of precipitation and atmospheric processes
as well as serve as a benchmark for existing models and might be useful
in e.g. drought and flooding hazard assessment. On a 
more general level, our analyses show that new insights can be
obtained from taking the very general point of view of complexity
and self-organization theory. It may serve as an 
example of how to use this approach in situations that seem too
complex to be accessible to quantitative analysis.

O.P. would like to thank G. Peters for his support. K.C. gratefully acknowledges
the financial support of U.K. EPSRC through grant nos. GR/R44683/01 and
GR/L95267/01. The MRR data collection was supported by the EU under
the Precipitation and Evaporation Project (PEP) in 
BALTEX.

$^*${To whom correspondence should be addressed,
E-mail: k.christensen@ic.ac.uk.}


\begin{thebibliography}{40}
\bibitem{1}
H.E. Hurst,
Nature {\bf 180}, 494 (1957).

\bibitem{2}
H.E. Hurst,
{\em Long-Term Storage: An Experimental Study}
(Constable \& Co. Ltd., London, 1965). 

\bibitem{3}
MMR, Physical Basis, pp. 21 (1998).
Available from METEK GmbH, Fritz-Stra{\ss}mann-Stra{\ss}e 4, D-25337 Elmshorn,
Germany.

\bibitem{4}
Information on the BALTEX project is available from the website 
http://w3.gkss.de/baltex/.

\bibitem{5}
D. Atlas, R.C. Srivastava, and R.S. Sekhon,
Rev. Geophys. Space Phys. {\bf 11}, 1 (1973).

\bibitem{6}
D. Klugmann, K. Heinsohn, and H.J. Kirtzel,
Contr. Atmos. Phys. {\bf 69}, 247 (1996).

\bibitem{7}
M. L\"{o}ffler-Mang and M. Kunz,
J. Atmos. Oceanic Technol. {\bf 16}, 379 (1999).

\bibitem{Mie1908}
G. Mie, Ann. Phys. {\bf 25}, 377 (1908).

\bibitem{8}
S. Lovejoy, Preprints, 20th Conf. On Radar Met., AMS, Boston 476 (1981).

\bibitem{9}
S. Lovejoy and B.B. Mandelbrot, Tellus 37A, 209 (1985).

\bibitem{10}
S. Lovejoy and D. Schertzer,
J. Appl. Meteorol. {\bf 29}, 1167 (1990).

\bibitem{11}
J. Feder, 
{\em Fractals} (Plenum Press, New York, 1988).

\bibitem{12}
P. Bak, C. Tang, and K. Wiesenfeld,
Phys. Rev. Lett. {\bf 59}, 381 (1987).

\bibitem{13}
P. Bak,
{\em How Nature Works: the science of self-organized criticality}
(Springer-Verlag, New York, 1996).

\bibitem{14}
H.J. Jensen, {\em Self-Organized Criticality: Emergent Complex
Behavior in Physical and Biological Systems} (Cambridge University Press, 1998).

\bibitem{15}
B. Gutenberg and C.F. Richter,
Bull. Seismol. Soc. Am. {\bf 34}, 185 (1994).

\bibitem{16}
V. Frette, K. Christensen, A. Malthe-S{\o}renssen, J. Feder, T. J{\o}ssang,
and P. Meakin,
Nature {\bf 379}, 49 (1996).
\end{thebibliography}
\end{document}